\documentclass[
prd,twocolumn,
,showpacs,amssymb,superscriptaddress,aps,nofootinbib
]{revtex4-2}
\usepackage{graphicx}
\usepackage{color}
\input{epsf}

\usepackage{amsmath,amssymb}
\usepackage{bm}
\usepackage{times}
\usepackage{ulem}
\usepackage[colorlinks=true]{hyperref}
\hypersetup{colorlinks=true,
            citecolor=blue,
            linkcolor=red,
            urlcolor=red}

\usepackage{amssymb,amsmath,amsfonts,mathrsfs,url}
\usepackage{xcolor}
\usepackage[colorlinks=true]{hyperref}
\hypersetup{citecolor=blue,linkcolor=red}
\usepackage{multirow,array}
\DeclareMathAlphabet{\pazocal}{OMS}{zplm}{m}{n}

\newcommand{\mM}{\text{Met}(M)}

\begin{document}
%%%%%%%%%%%%%%%%%%%%%%%%%%%%%%%%%%
%\twocolumngrid
% % %
\title{Geometric distances between closed universes}
\date{\today}
\author{Arthur G. Suvorov\thanks{arthur.suvorov@ua.es}}
\affiliation{Departament de F{\'i}sica Aplicada, Universitat d'Alacant, Ap. Correus 99, E-03080 Alacant, Spain}
\affiliation{Theoretical Astrophysics, Eberhard Karls University of T\"ubingen, T\"ubingen 72076, Germany}

\begin{abstract}
\noindent The large-scale structure of the Universe is well approximated by the Friedmann equations, parametrized by several energy densities which can be observationally inferred. A natural question to ask is: How different would the Universe be if these densities took on other values? While there are many ways this can be approached depending on interpretation and mathematical rigor, we attempt an answer by building a ``history space'' of different cosmologies. A metric is introduced after overcoming technical hurdles related to Lorentzian signature and infinite volume, at least for topologically closed cases. Geodesics connecting two points on the superspace are computed to express how distant---in a purely geometric sense---two universes are. Age can be treated as a free parameter in such an approach, leading to a more general mathematical construct relative to geometrodynamical configuration space. Connections to bubble inflation and phase transitions are explored.
\end{abstract}

\maketitle

%We find that {history}-space phase transitions can correspond to ``large'' distances, such as between Big Freeze and Bouncing universes, which could have implications for eternal inflation theory.
%\vspace{10pt}
%\begin{indented}

%\end{indented}
%\affiliation{}
% After overcoming some technical hurdles related to Lorentzian signature and infinite volumes, w
%%%%%%%%%%%%%%%%%%%%%%%%%%%%%%%%%%

%\noindent{\it Keywords}: Cosmology, Geometrodynamics, Superspace, Eternal inflation

%\maketitle

%\newpage

\section{Introduction}
A cornerstone of contemporary cosmology is the Friedmann–Lema{\^i}tre–Robertson–Walker (FLRW) metric. When plugged into the Einstein equations with a perfect fluid source, one obtains the Friedmann equations for a scale factor that depends on the particulars of the matter/energy content at large \cite{wein08}. These equations also can, and nowadays do despite Einstein's infamous blunder remark, feature a cosmological constant. Modern models typically consider four contributors, $\Omega_{i}$ \cite{koy16}. These represent radiation (of various flavours; $\Omega_{R}$), the sum of baryonic and dark matters ($\Omega_{M}$), spatial curvature (non-zero if the universe is not topologically flat; $\Omega_{k}$), and finally the cosmological constant (or vacuum more generally; $\Omega_{\Lambda}$). Appropriately normalized at some ``present time'', their dimensionless counterparts, $\Omega_{i,0}$, sum to one. One of these parameters is not independent therefore and so the large-scale structure of the emergent universe, including its ultimate fate, is thought to be encoded by just three independent numbers. 

At least within this simplified picture of a homogeneous and isotropic universe permeated by a smooth, cosmological fluid, a few different outcomes (``phases'') can await a persevering observer (see, e.g., Figure 5 in Ref.~\cite{ps03}). If, for example, the universe was closed with vanishing cosmological constant (so that $\Omega_{k,0} < 0$ but $\Omega_{\Lambda,0} = 0$) then expansion must eventually halt, ushering in an epoch of contraction culminating in a ``Big Crunch''. If instead $\Omega_{\Lambda,0} \gg \Omega_{M,0}$, the universe would be such that it was collapsing in the distant past, until contraction was halted by the repulsion established by a large, positive vacuum energy, since rebounding towards its presently expansionary state (``Big Bounce''). In a perhaps less dramatic but no less important way, the presence or absence of cosmic architecture ---  filaments, clusters, and voids --- is tied to the matter density, $\Omega_{M,0}$; the particulars can be studied with $N$-body simulations \cite{ang10,dragons}. By extension, what values these $\Omega_{i}$ take has implications for the possibility of life and the anthropic principle, shaping various habitability zones and typical proximities between star-planet systems \cite{barn22}.

Here we take a more mathematically direct approach to thinking about how universes can differ on large scales, following the geometrodynamics program of Wheeler \cite{wheeler3,wheeler4,giu09}. {In that context, one works with a \emph{configuration space} of initial data. Here, by contrast, we introduce mathematical machinery to work with the full four-dimensional spacetime via what could be called a ``history space''. There are some conceptual advantages of such a scheme, notably relating to phase transitions, which we highlight in Section~\ref{sec:history}.} Eventually, the cosmological arena described above can be represented as a set of distinct FLRW metrics. As can be proven rigorously \cite{clark10,ebin,gil1,gil2}, the collection of all \emph{Riemannian} metrics that can be placed on some manifold, $M$, can itself be viewed as an (infinite-dimensional) manifold, $\mM$. In this paper, we carve out a subspace of $\mM$ that corresponds only to different cosmological metrics. Some technical hurdles relating to (i) spacetime signature [i.e. $\mM$ is a manifold only for positive-definite metrics] and (ii) volume (universes which are not closed are of infinite volume) are addressed in the process, via the generalized Wick rotations introduced by Visser \cite{vis17} and by restricting our attention to closed universes ($\Omega_{k,0} < 0$), respectively.  

%relating to spacetime \emph{histories} rather than \emph{initial data}, as is more familiar from the geometrodynamics literature,
%This work is not intended to lay rigorous foundations for the theory of FLRW superspaces, but rather aims to construct some explicit examples of invariant distance measures between some interesting but standard cosmologies. 
%We find that such a space admits complicated curvature

This purpose of this work is to demonstrate, using established theory within differential geometry, that a simple and well-behaved quantification for a notion of \emph{distance} between universes can be formalized. Although primarily of mathematical interest, we speculate that geodesic distances within the FLRW superspace could describe separations between ``bubbles'' in the global picture of eternal inflation discussed by Guth and others since \cite{guth1,guth2,guth3}. 

This short paper is organised as follows. In Section \ref{sec:cosmology} we introduce some cosmological basics pertaining to FLRW metrics. Section \ref{sec:superspace} is devoted to the construction of superspaces generally, and then describes how we propose to extract metrics of Riemannian signature (Sec.~\ref{sec:tech1}) and finite volume (Sec.~\ref{sec:tech2}) from an FLRW family. Numerical results are given in Section \ref{sec:results}, with discussion provided in Section \ref{sec:disc}.

%he final space we work within is then built in Section \ref{sec:cossup}, w
%As it turns out, this set admits a rich geometric structure which allows for, amongst other things,  to be defined.   

%which are Einstein defines Wheeler's \emph{superspace}, and provides a means to quantify the relationship between different spacetime structures in a purely geometrical setting. In particular, geodesics on superspace can be used to `measure' a distance between distinct metrics.

\section{Cosmological setup} \label{sec:cosmology}

\begin{figure}
\begin{center}
\includegraphics[width=0.49\textwidth]{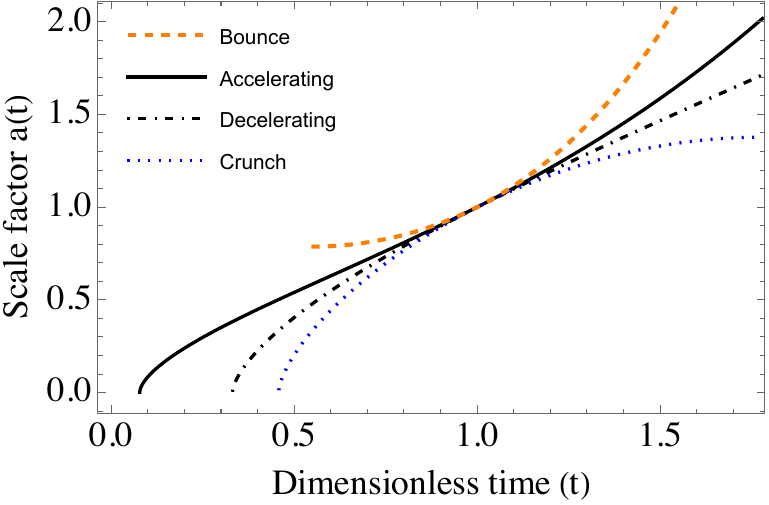}
\caption{Example solutions to the Friedmann equations \eqref{eq:friedmann}, corresponding to different ultimate fates as per the plot legends, with $\Omega_{i}$ as specified in the text. The curves continuously converge at $t=T=1$ due to the boundary conditions, though have different ages measured either from an initial Bang $(a=0)$ or present epoch (local minimum in $a$) in Bounce models (orange dashed curve). The Crunch model (dotted blue curve) reaches a local maximum at $t \approx 1.78$ and would eventually collapse $(a \to 0)$ had we extended the domain.
\label{fig:frwexampel}}
\end{center}
\end{figure}

Throughout we work with the standard family of FLRW line elements in four dimensions, geometrized units ($G = c = 1$), and reduced-circumference polar coordinates $\{t,r,\theta,\phi\}$,
\begin{equation}\label{eq:frw}
ds^2 = -dt^2 + a^2 \left( \frac{dr^2}{1 - kr^2} + r^2 d \theta^2 + r^2 \sin^2 \theta d \phi^2 \right),
\end{equation}
for (dimensionless) scale factor $a$ depending only on time, and curvature term $k$ {related to the spacetime topology and its volume in cases of closed universes ($k>0$)}. Associated with \eqref{eq:frw} are the Friedmann equations, 
\begin{equation}\label{eq:friedmann}
\begin{aligned}
\hspace{-0.05cm}\frac{1}{a}\frac{da}{dt}& = H_{0} \sqrt{ \Omega_{R} + \Omega_{M} + \Omega_{k} + \Omega_{\Lambda} } \\
&\equiv H_{0} \sqrt{ \Omega_{R,0} a^{-4} + \Omega_{M,0} a^{-3} + \Omega_{k,0} a^{-2} + \Omega_{\Lambda,0} },
\end{aligned}
\end{equation}
such that $a(T)=1$ for ``present'' (see below) time $T$. In cosmological studies it is typical to maintain the Hubble constant $H_{0}$; by appropriately scaling the units of length, on top of our geometric choices earlier, we can however set this to unity without loss of generality\footnote{For concreteness, if $H_{0} = 100 h \text{ km} \text{ s}^{-1} \text{ Mpc}^{-1}$ in physical units, then $H_{0} \approx 10^{-10} h \text{ ly}^{-1}$ in geometrized units, and thus we express lengths such that $\text{ly} \approx 10^{-10} h$. For $h \sim 0.7$, our normalized ``unit'' of time $\tilde{t}$ (see, e.g., the horizontal axis on Fig.~\ref{fig:frwexampel}) is thus $\sim 14$~Gyr in physical units. We, however, drop the overhead tilde for simplicity.}. Keeping $H_{0}$ within \eqref{eq:friedmann} does not alter the results or conclusions obtained in this work. {Having fixed $H_{0}$ and $a(T)$ however, the parameter $k$ must be left free so that $\Omega_{k,0}$ is not fixed; in general, we have the relation $\Omega_{k} = - k / a^2 H_{0}^2$ so that $\Omega_{k,0} = - k$.} Even so, note that $\sum \Omega_{i,0} = 1$. For the remainder of this work, we can thus fix
\begin{equation} \label{eq:fixedcond}
\Omega_{k,0} = 1 - \Omega_{\Lambda,0} - \Omega_{M,0} - \Omega_{R,0}.
\end{equation}
While this leads to triangular no-go regions where the spacetime is not closed (see Sec.~\ref{sec:tech2}), it is a familiar parameterisation and so we use it here. It should be emphasized that the above setup implies that the boundary value, $T$, does not represent an age, which is instead the difference between $T$ and some $t_{0}$ [$=t_{0}(\Omega_{i})$; permitted to be negative], which either corresponds to a singularity [$a(t_{0})=0$] or the beginning of the latest epoch (most recent local minimum of $a$) in Bouncing universes. Equation \eqref{eq:friedmann} can be integrated using standard methods for any given combination of $\Omega_{i}$. 

%within the global {history} space
The 2018 Planck data release indicates the following values are preferred for our universe \cite{planck,planck1} (confidence levels quoted at $1 \sigma$), assuming general relativity: $\Omega_{\Lambda} = 0.6847 \pm 0.0073$, $\Omega_{M} = 0.3153 \pm 0.0073$, and $\Omega_{k} = -0.0096 \pm 0.0061$; these imply $\Omega_{R} \approx - \Omega_{k} \approx 0.0096$. Note that $\Omega_{k} < 0$, indicating a closed universe, is consistent with the data. Figure~\ref{fig:frwexampel} shows an example set of solutions to Eq.~\eqref{eq:friedmann} corresponding to different phases. The solid, black curve corresponds to the parameters quoted above, indicating a universe which is expanding at an accelerating rate. Also shown are cases with the same $\Omega_{R,0} = 10^{-2}$ but various tuples $\{\Omega_{M},\Omega_{\Lambda}\}$:  $\{0.5,3\}$ (Bounce; orange dashed), $\{1, 0.1\}$ (decelerating expansion; black dotdashed), and $\{2,-0.5\}$ (Crunch; blue dotted), with $\Omega_{k,0}$ fixed by condition \eqref{eq:fixedcond}.

%(i.e. restricting $\alpha$ and $\beta$),

\section{Cosmological superspaces} \label{sec:superspace}

We now turn to the main goal of this work: understanding how one might assign numerical differences between different evolutionary tracks (such as those in Fig.~\ref{fig:frwexampel}). Again we emphasize that such an aim is primarily mathematical in nature, but notably \emph{reduces} the degrees of freedom traditionally present when considering differences between spacetimes. With a distance measure, a single, well-behaved scalar function can be used to quantify such differences, thereby skirting (often subtle) physical issues present in cosmology.

The set of all \emph{Riemannian} metrics over a ``well behaved'' (compact suffices; see Ref.~\cite{clark10}) \emph{Riemannian} manifold $M$ of dimension $n$, denoted $\mM$, admits a manifold structure \cite{gil1,gil2}. Points in $\mM$ are thus metrics on $M$: each $g \in \mM$ corresponds to a positive-definite, symmetric $(0,2)$-tensor over $M$. One may introduce\footnote{There are some mathematically attractive features of using expression \eqref{eq:gilmed} instead of another member from the more-general class of so-called ``alpha metrics'' (such as the DeWitt metric \cite{gil2}) which can be introduced on $\mM$; the interested reader can consult Ref.~\cite{gil1} for a discussion. Using a different metric from this class does not qualitatively change the picture discussed here.} a metric (together with Levi-Civita connection) on $\mM$ via a trace:
\begin{equation} \label{eq:gilmed}
G(\alpha,\beta) = \int_M d^{n} x \sqrt{g}\, \text{tr} \left( g^{-1} \alpha g^{-1} \beta \right) ,
\end{equation}
where $\alpha$ and $\beta$ are tangent vectors to the space of metrics at some ``reference point'' $g$, assuming that the integral converges. This rather general construction, though possibly relevant in studies of geometrodynamics and quantum gravity more broadly \cite{wheeler3,wheeler4,giu09}, is overkill for our purposes. We instead consider some appropriate restriction $\text{Met}_{\text{FLRW}}(M) \subset \mM$, which leads to a finite-dimensional and simpler superspace. Before doing so however, there are two technical obstacles that we must overcome: that related to signature (Sec.~\ref{sec:tech1}) and volume (Sec.~\ref{sec:tech2}). The reader interested only in the final result can skip to Sec.~\ref{sec:cossup}.

%As it stands, expression \eqref{eq:gilmed} is defined over the entirety of $\mM$, which is both numerically and conceptually challenging to work with.

\subsection{History space versus configuration space} \label{sec:history}

%In the gravitational sector, the ultraviolet regularity is made manifest by taming of the most prominent singularities of GR. In cosmology, the big bang and big crunch singulari- ties are replaced by a big bounce. (In fact, all strong curvature singularities are tamed [11], including the ‘big-rip’-type and ‘sudden death’-type that cosmologists often consider (see e.g., [7, 12]).) 
%he Universe naturally goes through a maximum of the curvature, after which the latter can only decrease; this is achieved with the scale factor passing through a minimal value, and hence a bounc

{We take a brief detour to provide some contextual motivation for this work. In particular, a well-known machinery that can already be used to assign distances between cosmologies is the geometrodynamic \emph{configuration space} corresponding to the set of initial data \cite{wheeler3,wheeler4,giu09}. In such a program, one introduces a foliation of spacetime into a set of codimension one leaves, $\{\Sigma_{t}\}$. One need only consider a single member from an equivalence class if the $\Sigma$ are diffeomorphic, so that the superspace metric \eqref{eq:gilmed} can be constructed with $n=3$ and reference metric identified with the spacetime metric projected onto some hyperslice. For a globally hyperbolic spacetime, a privileged initial-data slice (Cauchy surface) can be chosen. What we construct, by contrast, is a \emph{history space} with a metric involving integrals over the full four-dimensional spacetime manifold ($n=4$), appropriately rotated so as to become Riemannian (Sec.~\ref{sec:tech1}), which incorporates the evolution of the scale factor. There are advantages to this.}

{Firstly, there is a technical obstruction to building the configuration space in cases {with holes}. Comparing bangs and bounces, it is unclear if there is a suitable choice for a fixed leaf --- common to each family --- to integrate over in defining \eqref{eq:gilmed} (though see Refs.~\cite{kos18,floch20}). {This stems from the fact that a spacetime manifold $M$ cannot, by design, contain any singularities and thus $t=t_{0}$ is excluded \emph{a priori} for a bang. One could fix a time $t = t_{0} + \epsilon$ for $\epsilon >0$, though this carries a degree of arbitrariness. A bounce, by contrast, has no such restrictions, though it is also unclear what constitutes reasonable initial data for an eternal universe. Given interest in bounce cosmologies from quantum-gravity considerations where singularities are tamed (see Ref.~\cite{bp20} for a recent review), it is natural to ask whether such a distance can also be defined between universes of different phase. The history-space metric of Sec.~\ref{sec:cossup} avoids the slicing problem \emph{a priori} through a global construction: individual points do not contribute to the integral and thus details of how one deals with singularities or initial data are moot.}

{A second consideration relates to interpretations of geometric quantities and distances (see Sec.~\ref{sec:interp}). History space naturally includes an age, meaning it may be more general in terms of what can be compared. For example, some proposals for eternal inflation postulate that locally (in spacetime) there exists some transition probability of vacuum decay, where bubbles (possibly corresponding to different initial data) begin to expand and essentially represent individual universes. In that context, ages are important as concerns collision or nucleation probabilities between bubbles. That is, the volume of a bubble is determined by its age and its expansionary rate, only the latter of which is set by initial data. Since individual bubbles need not have the same age, it is advantageous to consider a formalism where time integrals are taken so that this aspect can be incorporated (see Sec.~\ref{sec:inflation}).} 
%as uncorrelated transitions can occur

{Finally, even in cases with fixed cutoffs or ages, one may ask whether the two formalisms provide the same quantifications for geometric entities. For instance, it is not obvious that distances between points in the neighbourhood of transitions should share the same patterns between configuration and history spaces as differences in $a(t)$ can eventually be large even if initial data are infinitesimally separated (see Fig.~\ref{fig:frwexampel}).} {To that extent, it is worthwhile to construct theoretically distinct ways to calculate geometric quantities of interest in cosmology generally (at least in cases where the two can be directly compared).}

%

%Provided that the integrals \eqref{eq:metten} exist, we can now proceed with superspace construction.

\subsection{Technical hurdle I: signature}
\label{sec:tech1}

It is customary to take a 3+1 split of the spacetime and adopt the Riemannian 3-space as the base manifold $M$ when building a superspace (e.g. \cite{suv20,suv21}). {However, our aim is to provide a framework to describe distances encapsulating cosmological histories}. To account for the full time domain, we must shift attention away from the Lorentzian and towards the \emph{Riemannian}.

%We have intentionally  as we intend we are going to take the entire manifold. Usually one takes a $3+1$ split, but in a cosmological spacetime this s less than desirable.
One way to achieve this is through a generalized Wick rotation. The basic procedure, as introduced by Visser \cite{vis17}, is as follows. Provided that $M$ admits a \emph{time-orientable} Lorentzian metric $g_{L}$, there exists an everywhere non-vanishing timelike vector. Taking one member from this class, $V$, we can define the (in general complex) metric 
\begin{equation} \label{eq:riem}
g_{\epsilon} = g_{L} + i \epsilon \frac{ V \otimes V} {g_{L} (V,V) },
\end{equation}
or, expressed locally,
\begin{equation}
 (g_{\epsilon})_{\mu \nu} = (g_{L})_{\mu \nu} + i \epsilon \frac{ V_{\mu} V_{\nu} } {(g_{L})^{\sigma \tau} V_{\sigma} V_{\tau}}.
\end{equation}
One now obtains a positive-definite metric through $g_{E} = g_{\epsilon \rightarrow 2 i}$.
In the case of the FLRW family \eqref{eq:frw} (i.e., our $g_{L}$), where the $dt dx^{\mu}$ dynamics are trivial, the simple choice $V_{\mu} = v_{0} \delta_{t \mu}$ for any $v_{0}$ is already satisfactory --- essentially just flipping the sign in front of $dt^2$. Though this procedure has some drawbacks related to non-uniqueness, it is adequate for our demonstrative purposes: we have a means to convert a family of cosmological line elements \eqref{eq:frw} into one with the correct signature \eqref{eq:riem} to proceed with the superspace construction. Although not shown, we have experimented with different choices of $V$ finding the comforting result that calculated quantities are largely unchanged (effectively through a conformal rescaling).

%1991 visser.
%\citealt{vis17}.

\subsection{Technical hurdle II: volume}
\label{sec:tech2}
%The existence of \eqref{eq:gilmed}, as written, requires $M$ to be of finite volume. 

A second issue hindering the construction of $G$ relates to the volume of the Riemannian bases $(M,g_{E})$: the integral defining \eqref{eq:gilmed} may not exist unless the (rotated) spacetime has finite volume. Many universes monotonically increase in volume with time, which poses a problem. In fact this is a well-known issue in the eternal inflation literature, being directly related to the ``measure problem''. It remains largely unresolved in that context, though one proposed solution is to introduce a cutoff for the scale factor. This allows for the introduction of a finite probability measure so that one can sensibly talk about the chances of ending up in our local ``pocket'' within some multiverse \cite{sim10}. (Some simple experimentation suggests that results do not depend much on the exact cutoff). {An alternative method to solve this problem could be to introduce a conformal factor such that the integral \eqref{eq:gilmed} converges, as was considered in Ref.~\cite{suv21} for black holes. This could carry the advantage that an epoch of contraction and expansion need not be selected for bounces.}

%In a study dealing with asymptotically flat black holes \cite{suv21}, this technical issue was dealt with by taking a compact submanifold corresponding to a radial extent larger than the Schwarzschild radius of some largest possible black hole. This worked well in that instance since the geometry is tame at large radii from the horizon, but for a spacetime with a conformal scaling of the 3-geometry this is not the case.

However, such a truncation will only aid with convergence in the time dimension, not the spatial ones\footnote{{Note the finiteness problem also exists in the configuration space construction for $k \leq 0$, where a compact subdomain is often used.}}. We therefore further restrict our attention in this work to closed universes, i.e., $k>0$ in expression~\eqref{eq:frw}. Considering evolution up to some (finite) present time ($t \leq T$), the volume of $M$ is bounded and the integral \eqref{eq:gilmed} exists using $g_{E}$. While these are fairly severe restrictions, they allow us to proceed directly to the main idea. It should be pointed out that the integrals we evaluate therefore do not explicitly contain information pertaining to end-states ($t>T$), but still do implicitly as fixing the $\Omega_{i}$ at $t=T$ is sufficient to encode the cosmological future, provided of course that the Friedmann equation \eqref{eq:friedmann} continues to hold. Moreover, $\Omega_{k} < 0$ is not strictly inconsistent with observations as described in Sec.~\ref{sec:cosmology}. %so that the construction is perhaps not of purely 

%e joint constraint with BAO measurements on spatial curvature is consistent with a flat universe

%consider closed universes.

%\twocolumngrid

%%Since we ignore the radiation density $\Omega_{R}$ we can fix, without loss of generality, $\Omega_{\Lambda} = 1 - \Omega_{M} - \Omega_{k}$. 
%With respect to this basis, the tensorial components of \eqref{eq:gilmed} read (see, e.g., Ref.~\cite{suv21})

%While one typically varies $\Omega_{M}$ and $\Omega_{\Lambda}$ in cosmological studies, we opt to keep $\Omega_{k}$ free instead since this allows us to maintain a rectangular grid of coordinates (avoiding a triangular ``no-go'' zone with $\Omega_{M} + \Omega_{\Lambda} \leq 1$, as we have fixed $k=1$). 

\subsection{A metric on the space of rotated, closed universes}
\label{sec:cossup}

We restrict our attention to a domain in which the argument of $G$ only accepts tangent vectors to the space of rotated \emph{FLRW metrics} \eqref{eq:riem}. More precisely, we consider a submanifold\footnote{Whether $\text{Met}_{\text{FLRW}}$ does in fact define a \emph{submanifold} rather than just a \emph{subset} depends on the properties of the cosmological family \eqref{eq:frw} and $V$, which we assume are sufficiently well-behaved. As shown below, the metric $G_{\rm FLRW}$ we end up with is positive-definite and everywhere regular.} $\text{Met}_{\text{FLRW}}(M) \subset \mM$ which inherits a metric, which we also call $G$ by abusing notation, from the parent space via pullback. Natural coordinates $\boldsymbol{\Omega} = \{\Omega_{M,0}, \Omega_{\Lambda,0}, \Omega_{R,0}\}$ are introduced on this \emph{three-dimensional} space (recalling Eq.~\ref{eq:fixedcond}), which is ultimately our object of interest.

In general, we may therefore write, on $\text{Met}_{\text{FLRW}}(M)$,
\begin{equation} \label{eq:metten}
G_{ij}(\boldsymbol{\Omega}) = \int_{M} d^{4} x \sqrt{g_{E}}  g_{E}^{nk} \frac { \partial (g_{E})_{mn}} {\partial \Omega^{i}}   g_{E}^{\ell m}  \frac { \partial (g_{E})_{\ell k}} {\partial \Omega^{j}} ,
\end{equation}
dropping the ``FLRW'' label. Evaluating \eqref{eq:metten}, one can compute the six independent components $\boldsymbol{G}$. For example, the entry in the ``$\Omega_{l}$-$\Omega_{m}$'' direction, using a cutoff at $t=T$ (or any other finite time), reads
\begin{equation} \label{eq:mm}
G_{\Omega_{l} \Omega_{m}}(\boldsymbol{\Omega}) = 3\int^{T}_{t_{0}(\Omega_{i})} \frac{dt \left(2 \pi^2 a^3\right) \left(\frac{\partial \log a}{\partial \Omega_{l}}\right) \left( \frac{\partial \log a}{\partial \Omega_{m}} \right)}{\left(\Omega_{\Lambda,0} + \Omega_{M,0} + \Omega_{R,0}-1\right)^{3/2}} , 
\end{equation}
{where we use the fact that the 3-volume of a closed universe with general $k$ is $2 \pi^2 a^3/k^{3/2}$ and $k = -\Omega_{k,0}$. Note that the denominator in \eqref{eq:mm} diverges (is imaginary) for a flat (open) universe, as expected based on considerations outlined in Sec.~\ref{sec:tech2}. An equivalent result would be found had we kept $H_{0}$ free (for instance) instead of $k$.} {We also remark that there is no fundamental reason the entire 3-space must be integrated over in defining \eqref{eq:metten}: if one wished to work with individual patches, this would be conceptually straightforward to implement by taking a spatial submanifold (provided, at least, the patches can be topologically identified). If such a spatial truncation is used, restriction to closed universes would also be unnecessary (Sec.~\ref{sec:tech2}).
}

Note that $G$ does not itself provide the measure we seek: it defines inner products of tangent vectors rather than distances between points. All of the geometric equipment of $\text{Met}_{\text{FLRW}}(M)$ can be defined from \eqref{eq:mm} though, including the Christoffel symbols $\boldsymbol{\Gamma}$. In particular, the distance between two metrics $p$ and $q$ over $M$ is given by the length of a geodesic $\gamma: [0,1] \mapsto \text{Met}_{\text{FLRW}}(M)$ connecting them, viz.
\begin{equation} \label{eq:distance}
d(p,q) = \int^{1}_{0} d \lambda \sqrt{ G_{ij} \frac {d \gamma^{i}} {d \lambda} \frac {d \gamma^{j}} {d \lambda} }
\end{equation}
such that
\begin{equation} \label{eq:geoeqn}
0 = \frac {d^2 \gamma^{i}} {d \lambda^2} + \Gamma^{i}_{j k} \frac {d \gamma^{j}} {d \lambda} \frac {d \gamma^{k}} {d \lambda},
\end{equation}
for some affine parameter $\lambda$, where $\boldsymbol{\gamma}(0) = \boldsymbol{\Omega}_{p}$ and $\boldsymbol{\gamma}(1) = \boldsymbol{\Omega}_{q}$.

\begin{figure}
\begin{center}
\includegraphics[width=0.499\textwidth]{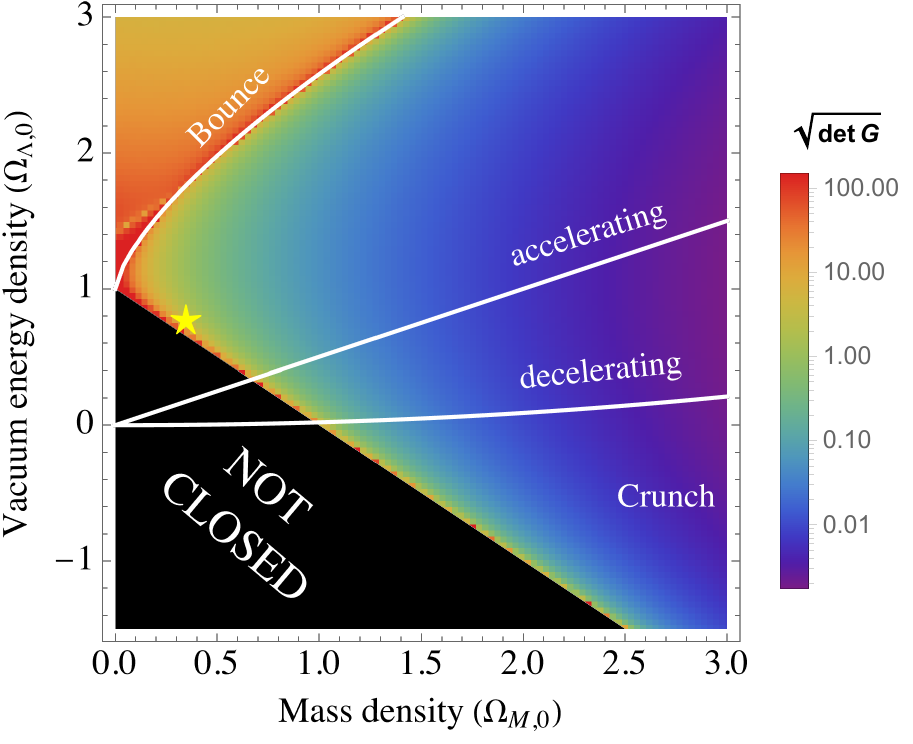}
\caption{Square root of the sub-determinant of the superspace metric $G_{\rm FLRW}$, numerically evaluated along the slice $\Omega_{R,0} = 10^{-3}$, as a function of $\Omega_{M}$ and $\Omega_{\Lambda}$; redder shades indicate a greater value of $\sqrt{\det G}$. Phase transitions are marked by white lines with overlaid text. The yellow star (approximately) marks the best-fit observational (Planck) values, as described in Sec.~\ref{sec:cosmology}. The coordinate region $\Omega_{\Lambda,0} \leq 1 - \Omega_{M,0} - \Omega_{R,0}$, corresponding to $\Omega_{k,0} \geq 0$, is excluded as the metric (i.e., expression \ref{eq:mm}) is not defined there.
\label{fig:detg}}
\end{center}
\end{figure}

\section{Results}
\label{sec:results}

We proceed by numerically solving the Friedmann equation \eqref{eq:friedmann}, similar to what was done in Sec.~\ref{sec:cosmology} and Fig.~\ref{fig:frwexampel} but over a fine grid of $\Omega_{M,0}, \Omega_{\Lambda,0}$, and $\Omega_{R,0}$ values to build an interpolating function $a(\Omega_{i,0},t)$ from which the components \eqref{eq:mm} can be accurately evaluated. It is convenient to fix an $\Omega_{R,0} = \text{constant} > 0$ slice as this allows us to consider two-dimensional diagrams in the $\Omega_{M,0}$--$\Omega_{\Lambda,0}$ subspace, as is more familiar from the literature (e.g. \cite{ps03}). We henceforth fix $\Omega_{R,0} = 10^{-3}$, though somewhat larger or smaller values do not affect the results much. In practice, the integral within \eqref{eq:mm} was evaluated using Simpson's rule with $N=2^{14}$ sampling points. Large $N$ are necessary as the integrand can be highly peaked, especially for coordinates neighbouring phase-transition boundaries; we anticipate errors at the $\sim 10^{-4}$ level, with some small extra systematics due to grid resolution, as below.

Taking a {$100 \times 100$ grid} of parameter values spanning the range $0 \leq \Omega_{M,0} \leq 3$ and $-3/2 \leq \Omega_{\Lambda,0} \leq 3$, we evaluate the necessary integrals to compute $\sqrt {\det G}$ along the slice $\Omega_{R,0} = 10^{-3}$; the results are shown in Figure~\ref{fig:detg}. Physically, this quantity weights the volume form and thus conveys a kind of averaged information about the local curvature.  As described initially in Ref.~\cite{isaac} and elsewhere since, one can deduce phase-transition boundaries by examining the existence or otherwise of roots of the right-hand side of equation \eqref{eq:friedmann}. These lines are marked in the Figure, as is the $\Omega_{k,0} \geq 0$ region which we exclude to ensure a closed universe (else $G$ is not defined). Note in particular that $\det G > 0$ everywhere {in the domain}, indicating a positive-definite (Riemannian) metric; {it does, however, diverge at the closed-flat boundary (though we set a maximum to the colour scale)}. A clear trend is seen that at larger values of $\Omega_{\Lambda,0}$ the determinant is larger, with $\det G$ reaching $\sim 10^{4}$ near the accelerating-Bounce boundary (sometimes called ``loitering'' universes), dropping by $\sim 8$ orders of magnitude when deep into parts of the parameter space corresponding to a Crunch. Curiously, near the observational values determined by Planck (yellow star) the metric $G$ is almost unimodular. %Such a range indicates that the curvature of the configuration space is highly non-trivial,

Equipped with $G$, evaluating the Christoffel symbols to then solve equation \eqref{eq:geoeqn} becomes a relatively simple numerical exercise.  Here we show some results, relating to geodesic paths (Sec.~\ref{sec:repex}) and distances more generally (Sec.~\ref{sec:distmaps}).

\subsection{Trajectories: a demonstration of curvature}
\label{sec:repex}
%(Here and below, the zero is dropped after the $i$ and $f$ subscripts to avoid clutter).

\begin{figure}
\begin{center}
\includegraphics[width=0.4\textwidth]{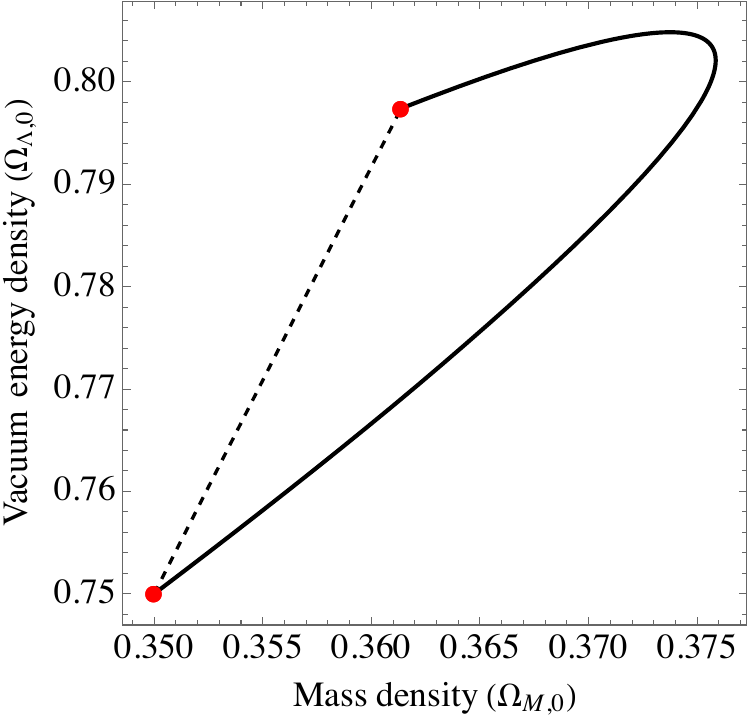}
\caption{Representative geodesic path (solid), on the FLRW superspace, between two Wick-rotated spacetimes (dots). The naive straight-line path is overlaid by a dashed curve. \label{fig:geosample}}
\end{center}
\end{figure}

To demonstrate the curvature of $\text{Met}_{\rm FLRW}$, we draw an example geodesic by solving equation \eqref{eq:geoeqn} between two points: $\Omega_{M,i}=0.35, \Omega_{\Lambda,i}=0.75$ (very roughly corresponding to our universe) and $\Omega_{M,f}=0.36, \Omega_{\Lambda,f}=0.80$. This endpoint still lies within the same state of accelerated expansion. The path is shown by the solid curve in Figure \ref{fig:geosample}, overlaid with a dashed, straight line (corresponding to a larger distance) for comparison. We find by direct integration that the geodesic curve corresponds to a (dimensionless) distance $d=0.108$ while the straight-line (SL) path has $d_{\rm SL} = 0.116$. Note these are a factor $\gtrsim 2$ larger than the Euclidean distance of the straight-line path, $d_{E} = 0.049$, illustrating both local and global curvature.

Consider instead now a geodesic connecting $\Omega_{M,i} = 1, \Omega_{\Lambda,i} = 2.65$ to the nearby point $\Omega_{M,f} = 1.004, \Omega_{\Lambda,f} = 2.53$; the path is shown in Figure~\ref{fig:geosample2}. The {former} coordinates correspond to a Bounce universe, as illustrated in Fig.~\ref{fig:detg}. {Despite the strong curvature present near the Bounce boundary, the geodesic is approximately straight though exhibits an overshooting and subsequent correction, similar to Fig.~\ref{fig:geosample}. We find in this case the comparable distances $d = 1.158$ and $d_{\rm SL} = 1.191$. It is interesting to note that despite the conceptually extreme nature of such a transition, the geometry of the {history} space remains smooth across this boundary (much like the event horizon of a Schwarzschild black hole viewed in Kruskal–Szekeres coordinates).} %Furthermore, such distances cannot be easily computed within the configuration space picture.
 
 %; a considerably larger difference than found previously, exceeding the Euclidean distance $d_{E}$ by a factor $\gtrsim 2$.
 %However, such a transition would seem to gravely violate causality, as the initial conditions are fundamentally distinct (Bang vs. Bounce; see Fig.~\ref{fig:frwexampel}). Even if $d$ is finite, we might therefore reject this possibility as unphysical, insofar as any transition is physical (though one is reminded of the ``horizon problem''). 

\begin{figure}
\begin{center}
\includegraphics[width=0.4\textwidth]{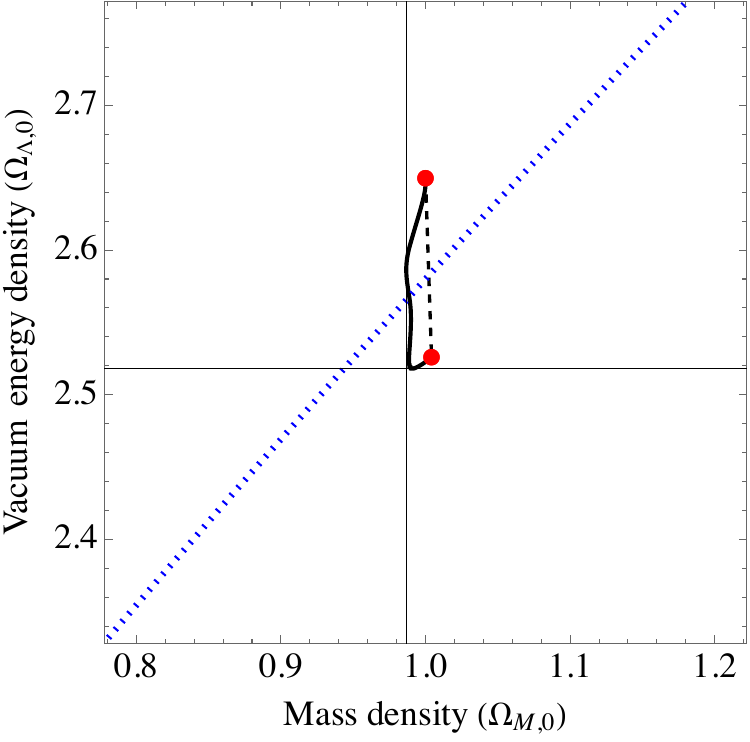}
\caption{Similar to Fig.~\ref{fig:geosample}, though instead depicting a geodesic crossing the boundary (dotted blue; see also Fig.~\ref{fig:detg}) separating an accelerating universe from a Bouncing one. \label{fig:geosample2}}
\end{center}
\end{figure}

\subsection{Distance maps}
\label{sec:distmaps}

\begin{figure*}
\begin{center}
\includegraphics[width=0.85\textwidth]{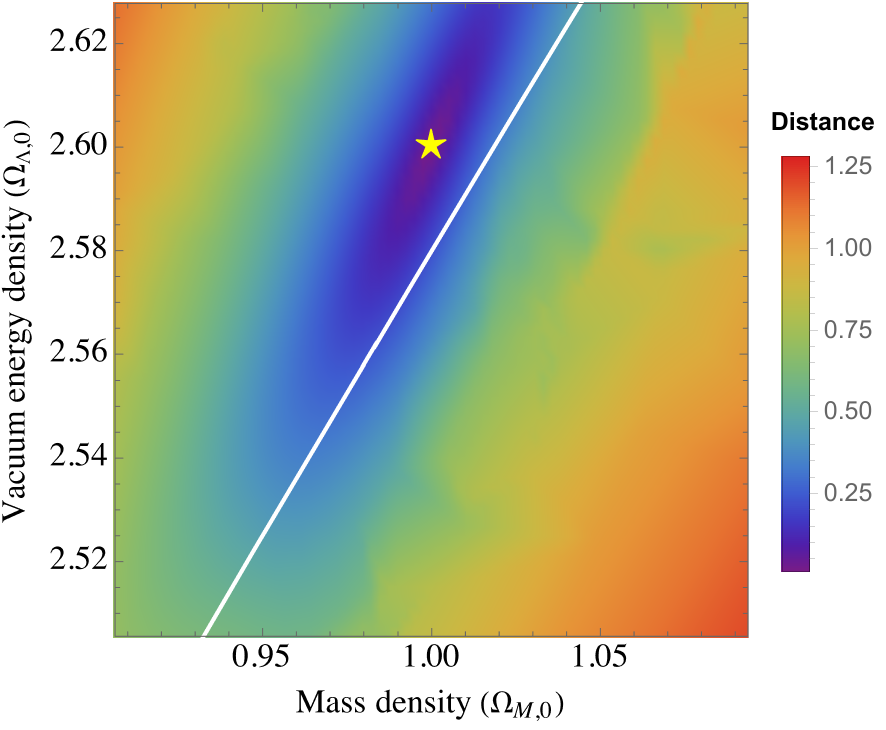}
\caption{A zoom in of the superspace distance map surrounding a universe with $\Omega_{M,i} = 1.0$ and $\Omega_{\Lambda,i} = 2.6$ (yellow star). Redder shades indicate greater (dimensionless) distances. The white line delineating the loitering boundary is overlaid, as in Fig.~\ref{fig:detg}. \label{fig:distmap}}
\end{center}
\end{figure*}

Instead of considering individual geodesics, we construct a suite of solutions corresponding to various end states\footnote{In practice, we use a shooting method to solve for geodesics by taking a rectangular grid of initial velocities, rather than enforcing boundary conditions.} to see how distances (in some appropriate units) behave on the {history} space more generally. {To highlight the value of a history-space approach, we present an example starting from a universe corresponding to a bounce with $\Omega_{M,i} = 1.0$ and $\Omega_{\Lambda,i} = 2.65$ in Figure~\ref{fig:distmap}}. Note we have zoomed in around the initial point and do not cover the entire submanifold. Nevertheless, moving in the negative $\Omega_{\Lambda,0}$ or positive $\Omega_{M,0}$ directions crosses into a different cosmological phase, namely that of accelerated expansion. %Note the irregularity of the graphed points comes about because 

{Despite the Euclidean distance being relatively large, heading diagonally in the direction parallel to the phase-boundary corresponds to small distances. That is, there is a preferred direction surrounding the line $\Omega_{\Lambda,0} \approx 2\Omega_{M,0} + 0.6$ such that the distance is in the neighbourhood of zero. While this line steepens for $\Omega_{M,0} \gtrsim 1.03$, it suggests that loitering universes are geometrically close to bounces, despite their conceptual differences. The existence of this line is tied to the fact that geodesics tend to trace contours of greater $\det G$, which is large near the boundary (see Fig.~\ref{fig:detg}). By contrast, heading deeper into the upper-left corner or towards universes with decelerating expansion corresponds to large distances.}

%This may appear to suggest that such a large starting value of $\Omega_{\Lambda,0}$ is disfavoured energetically, at least for a commensurately large starting value of $\Omega_{M,0}$. On the other hand, we see . This region roughly coincides with the contour $\det G \approx 1$ (see Fig.~\ref{fig:detg}). 

\section{Discussion}
\label{sec:disc}

In this Section, we offer some more speculative interpretations and discussion regarding the previous results and history-spaces more generally.

\subsection{Interpretations of paths and distances} \label{sec:interp}

{We can attempt an interpretation of the path depicted in Fig.~\ref{fig:geosample} (for example) by appealing to the principle of least action. As is well known, the Cauchy-Schwarz inequality ensures that any critical point of the energy functional is necessarily a constant-speed geodesic. Fig.~\ref{fig:geosample} therefore suggests that, postulating an energy-minimisation principle on the superspace, the path curves in a way that favours a blowout toward greater values of both the vacuum and matter energy densities. In the context of eternal inflation (see Sec.~\ref{sec:inflation}), this suggest that if the universe were to continuously transition (e.g. via nucleation) into a different state with $\Omega_{M,f}$ and $\Omega_{\Lambda,f}$ --- a different ``pocket'' with other values of these densities at small redshift --- it would do so in a way that is not monotonic with respect to the growth of the energy densities. This could lead to a kind of instability where, if history space paths could be connected to transitions through an action principle, runaways occur if the blowout becomes too large.}

{It is tempting to conjecture that if the universe were to transition to a new state (through some unspecified means), it would move along geodesics in order to remain ``nearby''. As depicted in Fig.~\ref{fig:distmap}, motions against $\det G$ correspond to the largest relative distances, where it is difficult to even shoot geodesics without high precision. On the other hand, the existence of a preferred direction along the diagonal lends indirect fortification to proposals involving minimal changes to the gravitational sector such that cosmological singularities are tempered \cite{bp20}: from a history-space perspective, geometrical structures even between some transitions are closeby. However, in the absence of any ``equations of motion'', it is not obvious that such an energy principle applies. More generally, if one were to postulate some torsion or non-metric connection or geodesic deviation on the history space, notions of distances would vary. It may be the case that \emph{any} type of phase transition, with the possible exception of an bounce-accelerating one if starting near the boundary, may be disfavoured on a history-space level. More general constructions will be investigated in future work.}

Despite uncertainties surrounding interpretation, having access to a single number which can be used to quantify differences between cosmology could prove practical in modified theories of gravity. Indeed, there was no \emph{a priori} need to restrict ourselves to the theory of general relativity or a four-dimensional spacetime. In other theories of gravity, the Friedmann equations \eqref{eq:friedmann} are typically modified in a way that includes an additional free parameter or several, but the mechanics of the calculation described in Sec.~\ref{sec:superspace} remain the same. The tools built here could thus also be used to speculate on how different a universe with the \emph{same} $\Omega_{i}$ but with stronger or weaker gravitational interaction would be. Owing to the unparalleled successes of general relativity, both theoretical and experimental, any theory of gravity hoping to supplant it ought to still be ``close'' to it, especially at energies far below the Planck scale. Another conjecture we might offer then is that only theories with distances less than some critical threshold could be considered viable. A similar idea was put forward in Ref.~\cite{suv21} for black holes: only those that are sufficiently close to the Kerr metric, consistent with gravitational-wave and X-ray measurements, should be considered astrophysical (see, e.g., Ref.~\cite{suv19} and references therein). Related ideas and concepts have been applied in studies of information theory and optimisation (e.g. \cite{w21}).

\subsection{Connections to bubble inflation} \label{sec:inflation}

%In this way, the metric $G$ will depend on a free parameter which is related to bubble age.
%(though this could be handled with a history space formalism by replacing the metric \eqref{eq:frw} by some bubble-filled metric)
Although what we call ``distance'' is something of an abstract quantity, it could represent something physical in the context of eternal inflation theory. If a given universe represented a patch within a multiverse --- divided into zones of different cosmological and/or fundamental parameters --- geodesic distance may be connected to the geometric structure of the overall arena (cf. Refs.~\cite{guth1,guth2,guth3}). {The reader should be aware however of a conceptual distinction between what may be considered the ``physical picture'', being that we do not deal with a single bubble-filled spacetime where many bubbles coexist. Instead we consider a set of individual bubbles of various size and age which coexist only as points on a history space. The advantage of this is we do not need to deal with matching/junction conditions on $M$, which can be difficult to impose except in some perturbative sense (see section 3 of Ref.~\cite{kleb} for a discussion).} {Thus far, we have considered cases where ages are fixed by the terminals of the integrals within \eqref{eq:mm}. We could, however, keep the upper terminal free so that the metric $G$ depends on a parameter related to an arbitrary bubble age, $\tau_{\rm bubble} \equiv T - t_{0}(\Omega_{i})$. Such a construction can be used to calculate a theoretical separation between two bubbles, one of which only just formed ($\tau_{\rm bubble} \sim 0$), to estimate their effective distance. Though lying beyond the scope here, combined with a probability measure for vacuum decay, a stochastic model for collisions could be formulated in a box of a given size.}

%and the same functions $a$
{To provide an example in this direction, consider a ``newborn'' cosmological development with $\Omega_{M,0} = 0.36, \Omega_{\Lambda,0} = 0.8$, as depicted in Fig,~\ref{fig:geosample}. We have $t_{0,i} = 0.043$ in dimensionless units, and we could thus take the integral \eqref{eq:mm} but with $T = t_{0,i}+\delta$ for some small $\delta>0$. A distance map surrounding such a cosmology is shown in Figure~\ref{fig:dist2} for $\delta = 10^{-2}$. The overall pattern of small distances along the parallel diagonal seen in Fig.~\ref{fig:distmap} persists, despite the small age and different phase (accelerating expansion). Note the slope of this short-distance curve is shallower however, viz. $\Omega_{\Lambda,0} \approx \Omega_{M,0} + 0.44$. We may anticipate therefore that the highest probability of bubble collision relates to neighbours with energy densities lying along diagonals of this kind, assuming an isotropic distribution for nucleation probabilities, 
the implications of which will be investigated elsewhere.} 

\begin{figure}
\begin{center}
\includegraphics[width=0.497\textwidth]{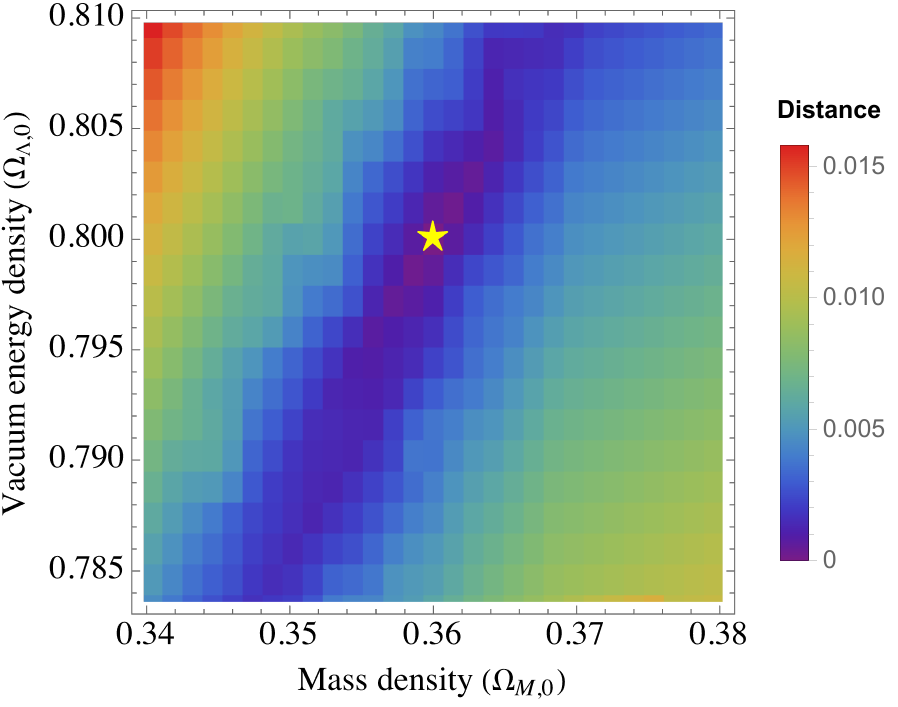}
\caption{Similar to Fig.~\ref{fig:distmap}, though for $\Omega_{M,0} = 0.36, \Omega_{\Lambda,0} = 0.8$ and relative to a newborn bubble with age $\tau_{\rm bubble}=0.01$. \label{fig:dist2}}
\end{center}
\end{figure}

%relative to a nearby `bubble' ($\Omega_{M,f} = 0.35, \Omega_{\Lambda,f} = 0.75$), as per the case in Fig.~\ref{fig:geosample}. The former has  while the latter has $t_{0,f} = 0.053$. We . 

%Based on a random walk from a starting point, position from the origin after some steps can be modelled. Using the fact that the expected distance from the origin after $N$ steps in a unit random-walk is $\sim \sqrt{N}$, for example, a typical horizon in the context of Fig.~\ref{fig:distmap} could be set after an appropriate normalization is employed. 

%The construction here can be used to assign probabilities in that context.

%Within Penrose's conformal cyclic cosmology paradigm \cite{ccc} distances, as considered here, between bouncing epochs could also invite a physical interpretation.

%which can be compared with their respective volumes.

%force $a(t) = 0$ up until some given bubble age $\tau_{\rm bubble}$, which will mean that the metric depends on this as a parameter. That is, replace the function $a$ by $\tilde{a}(t) = \Theta(t - \tau_{\rm bubble}) \left[ a(t) - a(\tau_{\rm bubble}) \right]$ where $\Theta$ is the Heaviside function and we require $a$ to be continuous. Such a construction can be used to calculate a separation between two bubbles of different ages which can be compared with their respective volumes.

\subsection{Conclusions and future directions} \label{sec:conclusions}

In this paper we describe a way to construct a single number as an answer to the question of how different two universes could be. Were an observer to ``move'' to a different cosmological-fluid spacetime where the various energy densities appearing in the Friedmann equations adjusted amongst themselves in such a way that the \emph{redshift zero} values $\Omega_{i,0}$ were different, we conjecture that they would do so in an energy-minimising way by following geodesic paths on the {history} space (see Figs.~\ref{fig:geosample} and \ref{fig:geosample2}). Such an approach suggests that universes without Bangs are somehow much more distant from our own than ones ending with a Big Crunch (see Fig.~\ref{fig:distmap}). %or that decelerate in their expansion.

%{It is worth pointing out that we have lumped all possible types of matter into a single parameter, $\Omega_{M}$. Small-scale aspects of universes with the same $\Omega_{M}$ but different proportions of dark and luminous behave in distinct ways, arguably the most fundamental of which stems from the fact that dark matter is crucial to forming overdensities prior to recombination. Without these strong gravitational wells in place, baryons cannot easily collapse en masse once photon pressure is alleviated. This leads to sparser clusters and likely the total dearth of galaxies altogether in the limit of no dark matter, even if $\Omega_{M}$ is large (though cf. Ref.~\cite{koy16}).}

In this context, we have restricted attention to large-scale structure and FLRW metrics \eqref{eq:frw}. However, the techniques could apply to cosmologies with more degrees of freedom provided a library of parameter combinations can be computed. A related avenue for extension involves the fundamental constants\footnote{Some theories of gravity suggest these values only masquerade as constants but are really dynamical fields (e.g. Brans-Dicke theory).}. While we have not considered them here, opting instead for geometric units and avoiding quantum altogether, we could also construct a superspace parameterised by Newton's constant, the speed of light, and even the fine structure constant $\alpha$ for some appropriate family of metrics. One could wax philosophical about the implications of such a space in the context of the anthropic principle. For instance, it seems necessary that $\alpha \approx 1/137$ else carbon atoms would be unstable as electrostatic repulsion between protons becomes overwhelming \cite{barrow03}. Such a fact could theoretically be reflected in the curvature of a superspace, with points (i.e. universes) such that $|\alpha_{f} - 1/137| \gg 0$ corresponding to insurmountable distances from others with $|\alpha_{i} - 1/137| \approx 0$. 

%Within an eternal inflation multiverse, ``anything that can happen will happen'' \cite{gv11}: but if one builds a probability measure from distance directly then meaningful, numerical answers can be given to event likelihood.

\section*{Acknowledgements}
I am grateful to Artem Tuntsov and Lucas Collodel for discussions. {I thank the anonymous referee for critical feedback, which significantly improved the quality of this manuscript.} Support provided by the Prometeo Excellence Programme grant CIPROM/2022/13 from the Generalitat Valenciana is gratefully acknowledged.

%%%%%%%%%%%%%%%%%%%%%%%%%%%%%%%%%%%%%%%%

\end{document}